\documentclass[journal]{IEEEtran}
\usepackage{makecell}
\usepackage{array}
\usepackage{graphicx,amssymb,amsmath}
\usepackage{multicol}
\usepackage[noadjust]{cite}
\usepackage{setspace}
\usepackage{subfigure}
\usepackage{graphicx}
\usepackage{float}
\usepackage {url}
\usepackage{stfloats}
\usepackage{amsthm,pifont}
\usepackage{flushend}
\usepackage{cases,subeqnarray}
\usepackage{bm,multirow,bigstrut}
\usepackage{amsmath, amsthm, amssymb}
\usepackage{textcomp}
\usepackage{latexsym,bm}
\usepackage{caption}
\usepackage{booktabs}
\usepackage{xcolor}
\usepackage{mathtools}
\usepackage{dsfont}
\usepackage{enumerate}
\usepackage{extarrows}
\usepackage{graphicx}                                                           
\usepackage{float} 
\usepackage{epsfig}
\usepackage{epsfig}
\usepackage{hhline}
\usepackage{epstopdf}
\usepackage[noend]{algpseudocode}
\newcounter{MYtempeqncnt}

\usepackage{cite}
\usepackage{bm}
\usepackage{cleveref}
\usepackage{multicol}       
\usepackage{multirow}       
\usepackage{array}          
\usepackage{colortbl}
\usepackage{makecell}
\definecolor{crimson}{RGB}{192,0,0}         
\definecolor{navy}{RGB}{47,85,151}         
\usepackage{bbding}
\usepackage{graphicx}

\usepackage{booktabs}
\makeatletter                               
\newif\if@restonecol
\makeatother

\makeatletter
\newif\if@restonecol
\makeatother

\usepackage[linesnumbered,ruled,vlined]{algorithm2e}
\usepackage{algpseudocode}
\theoremstyle{plain}

\theoremstyle{plain}

\usepackage{amsmath}

\IEEEoverridecommandlockouts
\begin{document}

\title{Jac-PCG Based Low-Complexity Precoding for Extremely Large-Scale MIMO Systems}
\author{Bokai Xu, Jiayi Zhang,~\IEEEmembership{Senior Member,~IEEE,} Jiaxun Li, Huahua Xiao,
and Bo Ai,~\IEEEmembership{Fellow,~IEEE}
\thanks{B. Xu, and J. Zhang are with the School of Electronic and Information Engineering, Beijing Jiaotong University, Beijing 100044, China. (e-mail: \{20251197,  jiayizhang\}@bjtu.edu.cn).}
\thanks{J. Li is with the Intelligent Game and Decision Lab (IGDL), Beijing 100044, China. (e-mail: lijiaxun@nuntil.edu.cn).}
\thanks{H. Xiao is with ZTE Corporation, State Key Laboratory of Mobile Network and Mobile Multimedia Technology, Shenzhen 518055, China.}
\thanks{B. Ai is with the State Key Laboratory of Rail Traffic Control and Safety, Beijing Jiaotong University, Beijing 100044, China.}
}
\maketitle

\begin{abstract}
 Extremely large-scale multiple-input-multiple-output (XL-MIMO) has been reviewed as a promising technology for future sixth-generation (6G) networks to achieve higher performance. In practice, various linear precoding schemes, such as zero-forcing (ZF) and regularized ZF (RZF) precoding, are sufficient to achieve near-optimal performance in traditional massive MIMO (mMIMO) systems. It is critical to note that in large-scale antenna arrays the operation of channel matrix inversion poses a significant computational challenge for these precoders. Therefore, we explore several iterative methods for determining the precoding matrix for XL-MIMO systems instead of direct matrix inversion.  Taking into account small- and large-scale fading as well as spatial correlation between antennas, we study their computational complexity and convergence rate. Furthermore, we propose the Jacobi-Preconditioning Conjugate Gradient (Jac-PCG) iterative inversion method, which enjoys a faster convergence speed than the CG method. Besides, the closed-form expression of spectral efficiency (SE) considering the interference between subarrays in downlink XL-MIMO systems is derived. In the numerical results, it is shown that the complexity given by the Jac-PCG algorithm has about $54 \%$ reduction than the traditional RZF algorithm at basically the same SE performance.    
		
	\end{abstract}

	\section{Introduction}
	 Looking forward beyond fifth-generation (B5G) and sixth-generation (6G), advanced novel wireless technologies have been developed to meet the increasingly demanding Internet service requirements \cite{[1], [19]}. One of these critical technologies is the extremely large-scale MIMO (XL-MIMO) by deploying an extremely large antenna array (ELAA) that can dramatically improve spectral efficiency (SE) \cite{[9],[4]}. With the number of antennas increasing, there are three significant challenges in XL-MIMO systems: computational complexity, scalability, and non-stationarities \cite{[3],[10]}. In particular, due to spatial non-stationary properties of channels, different parts may experience different characteristics \cite{[5]}. Proposed channel models have addressed these effects by considering the visibility regions (VRs) of users within multiple subarrays of the base station (BS) antenna array. 
	On the other hand, when the antenna array size becomes large, the channel directions between users become asymptotically orthogonal. To achieve near-optimal performance, linear precoders such as zero-forcing (ZF) and regularized ZF (RZF) are sufficient \cite{[20], [21]}. Additionally, they are widely used in engineering. Due to the large dimensions of XL-MIMO systems, matrix inversion becomes especially complex; direct inversion of an $K \times K$ matrix is typically of order \begin{math}
\mathcal{O}(K^{3}) 
\end{math}.\footnote{Note that several recursive matrix inversion algorithms have lower complexity than \begin{math}
\mathcal{O}(K^{3}) 
\end{math}. Due to the complicated recursive structure, these algorithms are difficult to implement in practice. The practical matrix inversion algorithms such as Gauss-Jordan elimination method have the complexity of \begin{math}
\mathcal{O}(K^{3}) 
\end{math}.} Therefore, it is necessary to develop more efficient matrix inversion methods, or approximations of them.
		
	To reduce the computational complexity of matrix inversion, several approaches have been proposed. In practical XL-MIMO systems, iterative methods are better candidates because they require less storage and are more computationally efficient than direct inversion methods such as singular value decomposition. As described in \cite{[15],[16]}, an expectation propagation (EP) detector with low complexity was proposed for XL-MIMO systems. Polynomial expansion (PE) was used to approximate the matrix inverse in each EP iteration in order to reduce the computational cost. Likewise, \cite{[14]} modeled the electromagnetic channel in the wavenumber domain using the Fourier plane wave representation for downlink multi-user communications. Moreover, they proposed the Neumann series expansion (NS) to replace matrix inversion with ZF precoding. However, matrix inversion is approximated by the summation of powers of matrix multiplications, which has a complexity order \begin{math}
\mathcal{O}(iK^{3}) 
\end{math} ($i$ is the iteration number) through the above methods. Additionally, increasing the number of iterations $i$ increases matrix inversion precision at the cost of greater computation. Our goal is to find precoding schemes that are as simple as possible.

	 A number of low complexity linear algorithms have been proposed in order to deal with crowded scenarios and to keep the cost of BSs as low as possible. These include \cite{[6], [17]}, to cite a few.
	 These works, however, do not consider non-stationary channels that appear when antenna arrays are scaled up, as is the case of XL-MIMO.     Moreover, some algorithms based on message passing \cite{[8]} and randomized Kaczmarz algorithm (rKA) \cite{[2],[7]} have been proposed to reduce the signal processing complexity.
	However, the convergence speed of the algorithm is plodding and requires a high number of iterations. 
	Several novel algorithms have been proposed, including mean-angle-based ZF (MZF), and tensor ZF (TZF) to reduce the complexity of the ZF \cite{[12]}. However, the performance of these algorithms in terms of SE still needs improvement. Therefore, there is a need for a new algorithm that can offer a better trade-off between performance and complexity. Nevertheless, to the best of our knowledge, there has yet to be any prior work investigating iterative matrix inversion methods for precoding in the context of XL-MIMO. 

	Motivated by the above observations, this correspondence aims to compare the iterative matrix inversion methods with the conventional RZF method, and its goal is to find the most competitive precoding scheme. The main contributions are given as follows:
	
	$\bullet$ Firstly, we apply Gauss-Seidel (GS), Jacobi Over-Relaxation (JOR) and Conjugate Gradient (CG) methods to circumvent direct matrix inversion considering the spatial non-stationary property. Besides, we analyze the advantages and disadvantages of several methods. Based on our analysis, the GS method has higher SE performance and faster convergence speed than the JOR method. Although the GS method performs better, it cannot be parallelized and has a high computational complexity. Both of them are not ideal matrices inversion algorithms.
		
	$\bullet$ Based on the traditional RZF precoding method, we propose an optimized scheme of CG, the Jacobi-Preconditioning CG (Jac-PCG) method, which has a faster convergence speed than the CG method. We analyze the SE  and bit error rate (BER) performance, considering the interference between subarrays. Simulation results \footnote{Simulation codes are provided to reproduce the results in this paper: https://github.com/BJTU-MIMO.} show that the Jac-PCG method achieves the RZF performance of direct matrix inversion with few iterations while hugely reducing the computational complexity. 

	{\emph {Organization:}} The rest of this paper is organized
	as follows. In Section \ref{sec:basic}, the XL-MIMO system model and non-stationary channel are introduced. In Section \ref{sec:methods}, we provide the details of designing iterative methods for matrix inversion. Numerical results are carried out in Section \ref{sec:simulation} and useful conclusions are drawn in Section \ref{sec:conclusion}.


\section{System Model}
\label{sec:basic}

\begin{figure}[t]
\includegraphics[width=3.3in]{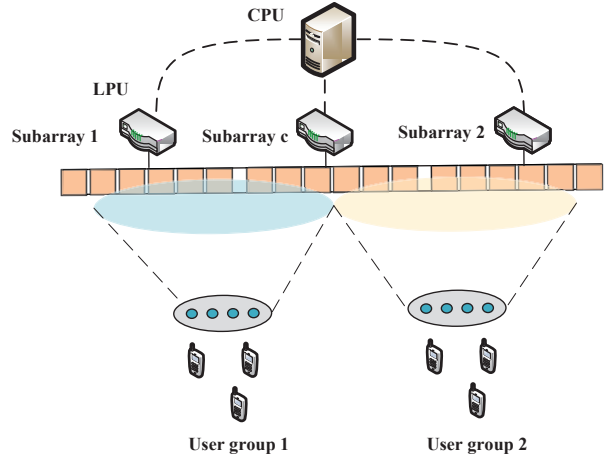}
\caption{Illustration of the investigated XL-MIMO system.}
\label{fig:system model}
\end{figure}

In this work, we examine the downlink transmission scenario of the XL-MIMO system, where the BS is equipped with a uniform linear array (ULA) consisting of $M$ antennas. $S$ denotes the number of fixed subarrays that splits an $M$ antenna array into disjoint groups of $M_{s}=M/S$ and each group has its own local processing unit for signal processing. Moreover, we divide $K$ single-antennas users into $L$ groups. So each user group has $K_{l}=K/L$ user equipments (UEs). As illustrated in Fig. \ref{fig:system model}, we assumed that $S=3, L=2$. The received complex signal
\begin{equation}
	{\bf y}={\bf H}^{H} {\bf x+n}
\end{equation}
with
\begin{equation}
	{\bf H}=\left[\begin{array}{cc}
{\bf H}_{1} & \bf{0} \\
{\bf H}_{c 1} & {\bf H}_{c 2} \\
\bf{0} & {\bf H}_{2}
\end{array}\right],
\end{equation} 
where ${\bf H}_{1}=\left[{\bf h}_{1,1}, {\bf h}_{1,2}, \cdot \cdot \cdot, {\bf h}_{1, K_{1}}\right] \in \mathbb{C}^{M_{1} \times K_{1}}$ is the channel matrix between the subarray $1$ and the  user group $1$. $ {\bf H}_{c}=[{\bf H}_{c 1}, {\bf H}_{c 2}]=\left[{\bf h}_{c1,1}, {\bf h}_{c 1,2},\cdot \cdot \cdot, {\bf h}_{c 1, K_{1}}, {\bf h}_{c 2,1}, {\bf h}_{c 2,2}, \cdot \cdot \cdot, {\bf h}_{c 2, K_{2}}\right] \in \mathbb{C}^{M_{c} \times K}$ is the channel matrix between the  subarray $c$ and all users. ${\bf H}_{2}=\left[{\bf h}_{2,1}, {\bf h}_{2,2},\cdot \cdot \cdot, {\bf h}_{2, K_{2}}\right] \in \mathbb{C}^{M_{2} \times K_{2}}$ is the channel matrix between the  subarray $2$ and the user group $2$. ${\bf x}\in \mathbb{C}^{M}$  contains the $M$ complex symbols messages, and ${\bf n} \in \mathbb{C}^{K} \sim  \mathcal{CN}(\mathbf{0}, \sigma^{2} \mathbf{I})$ is a white Gaussian noise vector. The $M_{s} \times 1$ channel vector with the channel coefficients of user $k$ to $M_{s}$ antennas of subarray $s$ is modeled as \cite{[11]} 
\begin{equation}
	\mathbf{h}_{s,k}=\sqrt{\mathbf{w}_{s,k}} \odot \overline{\mathbf{h}}_{s,k},
\end{equation}
where ${\bf w}_{s,k}$ gives the expression to large-scale fading effects; Path-loss is modeled as $\mathbf{w}_{s,k}=\Omega\left(\mathbf{d}_{s,k}\right)^{-\nu}$, where $\Omega$ is the path-loss attenuation coefficient. The distances between user $k$ and each antenna of the $s$-subarray  is defined as $\mathbf{d}_{s,k} \in \mathbb{R}^{M_{s}}$, and $\nu$ is the path-loss exponent. Channel effects resulting from small-scale fading are captured by by $\overline{\mathbf{h}}_{s,k} \sim \mathcal{N}_{\mathbb{C}}\left(\mathbf{0}, \boldsymbol{\Theta}_{s,k}\right)$, where $\boldsymbol{\Theta}_{s,k} \in \mathbb{C}^{M_{s} \times M_{s}}$ is the subarray channel covariance matrix that takes into account non-stationary and spatial channel correlation effects. The overall channel covariance matrix of the antenna array is $\boldsymbol{\Theta}_{k} \in \mathbb{C}^{M \times M}=\text{diag}\left(\boldsymbol{\Theta}_{1,k}, \ldots, \boldsymbol{\Theta}_{S,k}\right)$ and
\begin{equation}
	\boldsymbol{\Theta}_{k}=\mathbf{D}_{k}^{\frac{1}{2}} \mathbf{R}_{k} \mathbf{D}_{k}^{\frac{1}{2}},
\end{equation}
where $\mathbf{R}_{k} \in \mathbb{C}^{M \times M}$ is a symmetric positive semi-definite matrix that captures spatial channel correlation effects and $\mathbf{D}_{k} \in\{0,1\}^{M \times M}$ is a diagonal, indicator matrix  that models the non-stationary property of the channel through the VR concept.
We adopt the channel model described in \cite{[11]}, where each user is characterized by two properties of their VR: its center and length. The center of the VR is modeled as a uniform random variable, $c_{k}\sim \mathcal{U}(0, N)$, where $N$ represents the physical length of the XL-MIMO antenna array. And VR lengths are $l_{k}\sim \mathcal{LN}(\mu_{l},\sigma_{l})$.
When the linear precoding scheme such as RZF precoding is used, we can express $\bf x$ as 
\begin{equation}
	\bf x=Gs.
\end{equation}
It is similar to (2), the precoding matrix for $K$ users could be rewritten as a matrix form, i.e.,
\begin{equation}
	{\bf G}=\left[\begin{array}{cc}
{\bf G}_{1} & {\bf 0} \\
{\bf G}_{c 1} & {\bf G}_{c 2} \\
{\bf 0} & {\bf G}_{2}
\end{array}\right]
\end{equation}  
with
\begin{equation}
	\begin{array}{l}
{\bf G}_{1}=\beta_{1} {\bf H}_{1}\left({\bf H}_{1}^{H} {\bf H}_{1}+\xi {\bf I}_{K_{1}}\right)^{-1}=\beta_{1} {\bf H}_{1}{\bf P}_{1}^{-1}, \\
{\bf G}_{c}=\beta_{c} {\bf H}_{c}\left({\bf H}_{c}^{H} {\bf H}_{c}+\xi {\bf I}_{K}\right)^{-1}=\beta_{c} {\bf H}_{c}{\bf P}_{c}^{-1},\\
{\bf G}_{2}=\beta_{2} {\bf H}_{2}\left({\bf H}_{2}^{H} {\bf H}_{2}+\xi {\bf I}_{K_{2}}\right)^{-1}=\beta_{2} {\bf H}_{2}{\bf P}_{2}^{-1},
\end{array}
\end{equation}
where $\xi=\frac{1}{\mathrm{SNR}}=\frac{\sigma^{2}}{P}$ is the regularization factor. We define ${\bf P}_{i}={\bf H}_{i}^{H}{\bf H}_{i}+\xi {\bf I}$, ${\bf F}_{i}={\bf H}_{i}({\bf H}_{i}^{H}{\bf H}_{i}+\xi {\bf I}  )^{-1}$, $i\in \left \{ 1,2,c \right \}, $ and $\beta=\sqrt{P / \text{tr}\left({\bf F}_{i}^{H} {\bf F}_{i}\right)} >0$ represents the power control factor. Besides, ${\bf G}_{1}=[{\bf g}_{1,1}, {\bf g}_{1,2}, \cdot \cdot \cdot, {\bf g}_{1, K_{1}}] \in \mathbb{C}^{M_{1} \times K_{1}}$, ${\bf G}_{2}=\left[{\bf g}_{2,1}, {\bf g}_{2,2},\cdot \cdot \cdot, {\bf g}_{2, K_{2}}\right] \in \mathbb{C}^{M_{2} \times K_{2}}$, ${\bf G}_{c}=[{\bf G}_{c1}, {\bf G}_{c2}]=\left[{\bf g}_{c 1,1}, {\bf g}_{c 1,2},\cdot \cdot \cdot, {\bf g}_{c 1, K_{1}}, {\bf g}_{c 2,1}, {\bf g}_{c 2,2},\cdot \cdot \cdot, {\bf g}_{c 2, K_{2}}\right] \in \mathbb{C}^{M_{c} \times K}$  
Consequently, the received signal in (1) can be expressed as (8), where $s_{i,j}$ denotes the data symbol sent to the $j$-th user of the user group $i$, $n_{i,k} \sim  \mathcal{CN}(0, \sigma^{2})$. So the signal-to-interference-plus-noise ratio (SINR) of the $k$-th user of the user group $i$ can be expressed as (9). The total sum achievable data rate for all the UEs can be denoted by
\addtocounter{equation}{2}
\begin{equation}
	R_{\text {sum }}=\sum_{i=1}^{L} \sum_{k=1}^{K_{i}} \mathbb{E}\left\{\log _{2}\left(1+\gamma_{i, k}\right)\right\}.
\end{equation}
%
%
%

\begin{figure*}[!b]
\normalsize
\setcounter{MYtempeqncnt}{\value{equation}}
\setcounter{equation}{7}
\begin{equation}
\label{eqn_dbl_x}
y_{i, k}=\left({\bf h}_{i, k}^{H} {\bf g}_{i, k}+{\bf h}_{c i, k}^{H} {\bf g}_{c i, k}\right) s_{i, k}+\sum_{j=1, j \neq k}^{K_{1}}\left({\bf h}_{i, k}^{H} {\bf g}_{i, j}+{\bf h}_{c i, k}^{H} {\bf g}_{c i, j}\right) s_{i, j}+\sum_{j=1, m \neq i}^{K_{2}} {\bf h}_{c i, k}^{H} {\bf g}_{c m} s_{m, j}+n_{i, k}, k=1,2, \ldots, K_{i}
\end{equation}
\hrulefill
\begin{equation}
\label{eqn_dbl_y}
\gamma_{i, k}=\frac{\left|{\bf h}_{i, k}^{H} {\bf g}_{i, k}+{\bf h}_{c i, k}^{H} {\bf g}_{c i, k}\right|^{2}}{\sum_{j=1, j \neq k}^{K_{i}}\left|{\bf h}_{i, k}^{H} {\bf g}_{i, j}+{\bf h}_{c i, k}^{H} {\bf g}_{c i, j}\right|^{2}+\sum_{j=1, m \neq i}^{K_{m}}\left|{\bf h}_{c i, k}^{H} {\bf g}_{c m}\right|^{2}+\sigma^{2}}, k=1,2, \ldots, K_{i}
\end{equation}
\setcounter{equation}{\value{MYtempeqncnt}}
\hrulefill
\vspace*{4pt}
\end{figure*}

\section{Iterative Methods For Matrix Inversion}
\label{sec:methods}
As shown in (7), the RZF precoding requires the matrix inversion of large size, so its complexity 
\begin{math}
\mathcal{O}(K^{3})
\end{math} 
rises rapidly as the dimension of XL-MIMO expands. Some low-complexity processing schemes are proposed to overcome this problem. There are stationary and non-stationary methods for solving systems of linear equations, depending on whether the solution is expressed as finding the stationary point of some fixed-point iteration \cite{[18]}. 

We consider a SLE $\bf Ax=b$, where ${\bf A}\in \mathbb{C}^{m \times n}$, ${\bf b}\in \mathbb{C}^{m}$. The simple form of stationary iterative methods can be expressed as follows:
\begin{equation}
	\mathbf{x}^{(k)}={\bf B} \mathbf{x}^{(k-1)}+{\bf c},
\end{equation}
where neither $\bf B$ nor $\bf c$ depends upon the iteration count $k$. Non-stationary iterative methods are also called Krylov subspace methods. By minimizing the residual over the subspace formed, approximations to the solution are formed. Unlike equation (11), we get 
\begin{equation}
	\mathbf{x}^{(k+1)}=\mathbf{x}^{(k)}+\alpha_{k} \mathbf{z}^{(k)},
\end{equation}
and 
\begin{equation}
\mathbf{r}^{(k+1)}=\mathbf{b}-\mathbf{A} \mathbf{x}^{(k+1)}=\mathbf{r}^{(k)}-\alpha_{k} \mathbf{A} \mathbf{z}^{(k)}.
\end{equation}
A non-stationary iterative method involves the following operations at every step:
\begin{equation}
\begin{aligned}
	&\text{solve the linear system} \quad {\bf P}{\bf z}^{(k)}=\mathbf{r}^{(k)} ,\\ &\text{compute the acceleration parameter}\quad  \alpha_{k} , \\&\text{update the solution} \quad \mathbf{x}^{(k+1)}=\mathbf{x}^{(k)}+\alpha_{k} \mathbf{z}^{(k)} , \\&\text{update the residual} \quad \mathbf{r}^{(k+1)}=\mathbf{r}^{(k)}-\alpha_{k} {\bf A} \mathbf{z}^{(k)}.
	\end{aligned}
\end{equation}
In summary, the main difference between stationary and non-stationary iterative methods lies in the iteration scheme. Stationary methods maintain a fixed scheme throughout the iterations, while non-stationary methods adapt and change the scheme during the solution process.
\subsection{Stationary Iterative Methods}
Considering (5) and (7), we can rewrite the transmitted signal vector ${\bf x}_{i}$ as 
\begin{equation}
	{\bf x}_{i}=\beta_{i} {\bf H}_{i} \mathbf{P}^{-1}_{i} \mathbf{s}_{i}=\beta \mathbf{H}_{i}  {\bf w}_{i},  
\end{equation}
where ${\bf  w}_{i}={\bf P}_{i}^{-1} {\bf s}_{i}$ and $i\in \left \{ 1,2,c \right \}$.
 Equivalently, we have
\begin{equation}
	{\bf P}_{i}{\bf w}_{i}={\bf s}_{i}.
\end{equation}
We start with an examination of GS and JOR, being two well-known methods for solving a set of linear equations. Both methods solve (16) by iteratively calculating ${\bf w}_{i}^{t}={\bf A}_{i}{\bf w}_{i}^{t}+{\bf b}_{i}$. According to the GS method, ${\bf A}_{i}=-({\bf D}_{i}+{\bf L}_{i})^{-1}{\bf U}_{i}$, ${\bf b}_{i}=({\bf D}_{i} +{\bf L}_{i})^{-1}{\bf s}_{i}$, and the JOR method, ${\bf A}_{i}={\bf I}-\omega {\bf D}_{i}^{-1}{\bf P}_{i}$, ${\bf b}_{i}=\omega {\bf D}_{i}^{-1}{\bf s}_{i}$.  ${\bf D}_{i}$ contains the diagonal part of ${\bf P}_{i}$ and ${\bf L}_{i}$ and ${\bf U}_{i}$ are the strictly lower and strictly upper triangular parts of ${\bf P}_{i}$.

\subsection{Non-Stationary Iterative Methods}
In the context of massive MIMO (M-MIMO) signal processing schemes, the CG algorithm is a non-stationary iterative method for solving the matrix inversion problem. However, when the condition number of Hermitian positive definite matrix
\begin{equation}
	\operatorname{cond}({\bf P}_{i})=\|{\bf P}_{i}\|\left\|{\bf P}_{i}^{-1}\right\|
\end{equation}
 is large, the number of iterations of CG algorithm will significantly increase and the convergence speed will significantly slow down. Therefore, the PCG method is used to preprocess the coefficient matrix to achieve the purpose of reducing the number of iterations and accelerating the convergence speed.

The premise of utilizing PCG method is that the matrix ${\bf P}_{i}$ should be Hermitian positive definite. Firstly, due to the fact that ${\bf P}_{i}={\bf H}_{i}^{H}{\bf H}_{i}+\xi {\bf I}$ according to the definition, the matrix ${\bf P}_{i}$ is clear to be Hermitian. Secondly, by denoting an arbitrary nonzero vector as $\bf q$, we have
\begin{equation}
	{\bf q}{\bf P}_{i}{\bf q}^{H}={\bf q}({\bf H}_{i}^{H}{\bf H}_{i}+\xi {\bf I}){\bf q}^{H}={\bf q}{\bf H}_{i}({\bf q}{\bf H}_{i})^{H}+\xi {\bf q}{\bf q}^{H} .
\end{equation}
In XL-MIMO channels, where the channel matrix ${\bf H}_{i}$ is a full-rank matrix, ${\bf q}{\bf H}_{i}$ equals to a zero vector only when ${\bf q}$ is a zero vector. So we have ${\bf q}{\bf H}_{i}({\bf q}{\bf H}_{i})^{H}+\xi {\bf q}{\bf q}^{H} >0$ for all nonzero vectors, which indicates that ${\bf P}_{i}$ is positive definite.

%
%
%
%
%
%
%

Compared with the traditional CG algorithm, the preconditioning matrix ${\bf C}_{i}$ is defined in advance for the PCG algorithm. In XL-MIMO systems, where the interactions between distant elements can be considered negligible, leading to many zero elements in the matrix. Thus, the matrix ${\bf P}_{i}$ is a sparse matrix.
Basically, the PCG algorithm to solve (16) in the following four steps:

\begin{enumerate}[\underline {\bf Step 1}:]
\item Select the initial solution ${\bf w}_{i}^{0}$ and set the allowable error range as $\varepsilon > 0$.	
\end{enumerate}
\begin{enumerate}[\underline {\bf Step 2}:]
\item Calculate negative gradient direction ${\bf g}_{i}^{t}=\nabla f({\bf w}_{i}^{t})$, if $\|{\bf g}_{i}^{t}\|\leq\varepsilon$, the algorithm is terminated; else go to $\text{\underline {\bf step 3}}$.
	
\end{enumerate}
\begin{enumerate}[\underline {\bf Step 3}:]
\item Calculate
\begin{equation}
	{\bf w}_{i}^{t+1}={\bf w}_{i}^{t}+\lambda_{i}^{t}{\bf d}_{i}^{t},
\end{equation}
where $\lambda_{i}^{t}=\frac{({\bf g}_{i}^{t})^{H}{\bf z}_{i}^{t}}{({\bf d}_{i}^{t})^{H}{\bf P}_{i}^{t}{\bf d}_{i}^{t}}$, ${\bf z}_{i}^{t}={\bf C}^{-1}_{i}{\bf g}_{i}^{t}$. Calculate
\begin{equation}
	{\bf d}_{i}^{t}=-{\bf g}_{i}^{t}+\beta_{i}^{t-1}{\bf d}_{i}^{t-1},
\end{equation}
where $\beta_{i}^{t-1}=\frac{({\bf z}_{i}^{t-1})^{H} {\bf P}_{i}^{t} {\bf g}_{i}^{t}}{({\bf z}_{i}^{t-1})^{H} {\bf P}_{i}^{t} {\bf d}_{i}^{t}}$.
\end{enumerate}

\begin{enumerate}[\underline {\bf Step 4}:]
\item We set $t=t+1$.
\end{enumerate}

For the Jac-PCG method, We can get the preconditioning matrix ${\bf C}_{i}$ in the following way.	
Decompose ${\bf P}_{i}$ as 
\begin{equation}
	{\bf P}_{i}={\bf D}_{i}+{\bf L}_{i}+{\bf L}_{i}^{H},
\end{equation}
where ${\bf D}_{i}$ is the diagonal part of the Hermitian positive definite matrix ${\bf P}_{i}$ and ${\bf L}_{i}$ is the strictly lower triangular part of the Hermitian positive definite matrix ${\bf P}_{i}$. According to the Jacobi matrix splitting method, the preconditioned matrix ${\bf C}_{i}$ is denoted by ${\bf D}_{i}$. We summarize the main steps in Algorithm \ref{algorithm}. 
 


\section{Simulation Results}
\label{sec:simulation}

In this section, we compare the performance between the stationary and non-stationary iterative methods based on the RZF algorithm for the downlink precoding techniques in terms of SE and computational complexity. The simulation parameters are disposed in Table \ref{tab3}. The UEs are uniformly distributed in a square-cell area with a minimum distance of $30$ m to the BS. The extremely large antenna array is arranged in a uniform linear array (ULA) configuration, with a spacing of $2\lambda $ m between each antenna.\footnote{We can approximately ignore the effect of antenna mutual coupling, and all the users are in the near-field region because of the extremely large antenna array. Moreover, the gap between the investigated methods is more obvious when the antenna distance is set to $2\lambda$.}

\begin{figure}[t]
\centering
\includegraphics[width=3.2in]{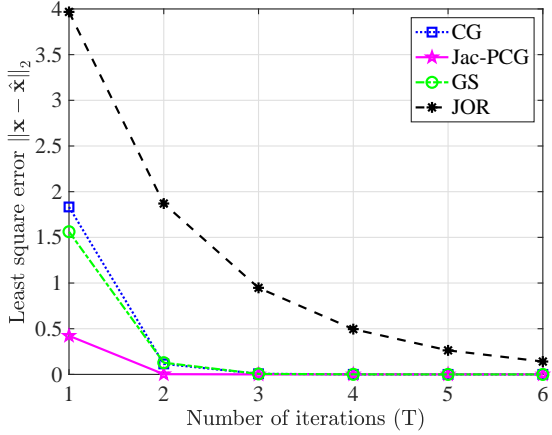}
\caption{Least square error of GS, JOR, CG, and Jac-PCG methods against the number of iterations.}
\label{fig:convergence}
\end{figure}

In Fig. \ref{fig:convergence}, we consider the least square error of the traditional RZF, CG, Jac-PCG, GS, and JOR methods against the number of iterations. It can be seen that the Jac-PCG algorithm has a faster convergence than the CG algorithm. When the number of iterations $T$ reaches $5$, all the methods can basically achieve convergence.

\begin{figure}[t]
\centering
\includegraphics[width=3.2in]{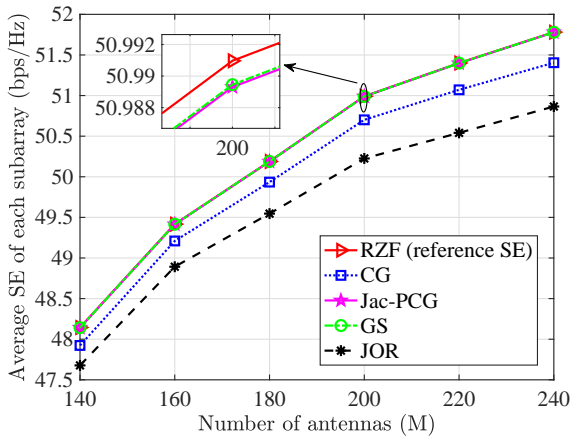}
\caption{Average DL SE as a function of the number of antennas for different matrix inversion schemes over non-stationary channels in XL-MIMO systems.}
\label{fig:SE_M}
\end{figure}

Fig. \ref{fig:SE_M} shows that the SE with the traditional RZF, CG, Jac-PCG, GS, and JOR methods against $M$ for non-stationary channels with $K_{l}=16$, $S=3$ and $T=5$. The SE achieved using RZF of direct matrix inversion serves as a higher bound and benchmark for the performance of other methods. It is obvious that with the number of antennas increases, the gap between the methods gradually increases. 
\begin{algorithm}[t]
		\caption{Low-Complexity Precoding Scheme Designed by the Jac-PCG Algorithm}
		\label{algorithm}
		\KwIn {The Hermitian matrix ${\bf P}_{i}$, and the number of the algorithm iterations $T$;}
		{\bf Initiation:} The iterative initial value ${\bf w}_{i}^{t}={\bf w}_{i}^{0}$, the initial residual error ${\bf r}_{i}^{t}=({\bf C}_{i}^{t})^{-1} {\bf s}_{i}^{t}$, the gradient direction ${\bf m}_{i}^{0}={\bf r}_{i}^{0}$;\\
		\For{$t=1,2,...,T$}
		{
			$\alpha_{i}^{t}=\frac{({\bf r}_{i}^{t})^{H} {\bf r}_{i}^{t}}{({\bf m}_{i}^{t})^{H} ({\bf C}_{i}^{t})^{-1} {\bf P}_{i}^{t} {\bf m}_{i}^{t}}$;
			
			Update ${\bf w}_{i}^{t+1}={\bf w}_{i}^{t}+\alpha_{i}^{t} {\bf m}_{i}^{t}$;
			
			Update ${\bf r}_{i}^{t+1}={\bf r}_{i}^{t}-\alpha_{i}^{t}({\bf C}_{i}^{t})^{-1} {\bf P}_{i}^{t} {\bf m}_{i}^{t}$;
			
			${\bf b}_{i}^{t}=\frac{({\bf r}_{i}^{t+1})^{H} {\bf r}_{i}^{t+1}}{({\bf r}_{i}^{t})^{H} {\bf r}_{i}^{t}}$;
			
			Update ${\bf m}_{i}^{t+1}={\bf r}_{i}^{t+1}+{\bf b}_{i}^{t} {\bf m}_{i}^{t}$;
		}
		
		\KwOut{${\bf w}_{i}^{t}$; }
		
	\end{algorithm}

\begin{table}[t]
\fontsize{9}{12}\selectfont
\setlength\tabcolsep{3pt} 
\begin{center}
\caption{Simulation Parameters in XL-MIMO Systems.}
\label{tab3}
\begin{tabular}{c | c | c|c}
\hline

{\bf Parameter} & {\bf Value}& {\bf Parameter}&{\bf Value}\\

\hline
\hline
Cell area&  $0.1 \times  0.1$ km$^{2}$  &$N$&$23.0610$ m\\
\hline
Min. distance&$30$ m &$K$&$32$\\ 
\hline
Array type& ULA& $S$&$3$\\
\hline
Antenna spacing&2$\lambda $ m&$l_{k}$&$\mathcal{LN}(0.1N,0.1)$ \\
\hline 
Carrier frequency&$2.6$ GHz&$c_{k}$&$\mathcal{U}(0,N)$\\
\hline 
  $\nu $&$3$& $\sigma ^{2}$&$-50$ dBm\\
\hline 
$T$ &5&$\Omega $&$4$\\
\hline 
\end{tabular}
\end{center}
\end{table}

\begin{figure}[t]
\centering
\includegraphics[width=3.2in]{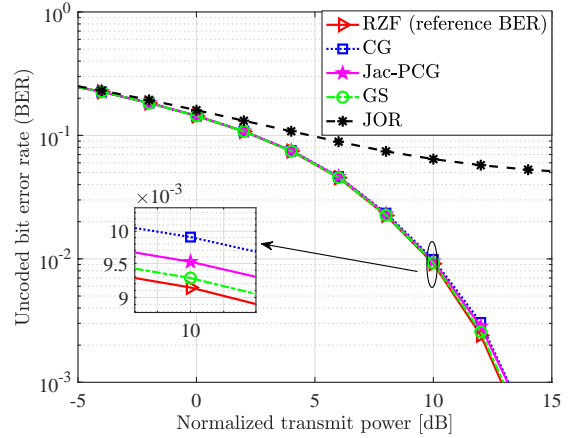}
\caption{BER performance of RZF, GS, JOR, CG and Jac-PCG methods against the normalized transmit power.}
\label{fig:BER}
\end{figure}

Fig. \ref{fig:BER} shows that the BER performance of the traditional RZF, CG, Jac-PCG, GS, and JOR methods vs. normalized transmit power for non-stationary channels in XL-MIMO systems. When the normalized transmit power is $10$ dB, we notice that the Jac-PCG method has $4.04 \%$ improvement than the CG method. In contrast, due to the slow convergence speed of the JOR method, when the number of iterations is $5$, its error is enormous compared with the actual value of the matrix inverse. Therefore, the JOR method is unsuitable for XL-MIMO systems because of its high BER.

\begin{figure}[t]
\centering
\includegraphics[width=3.2in]{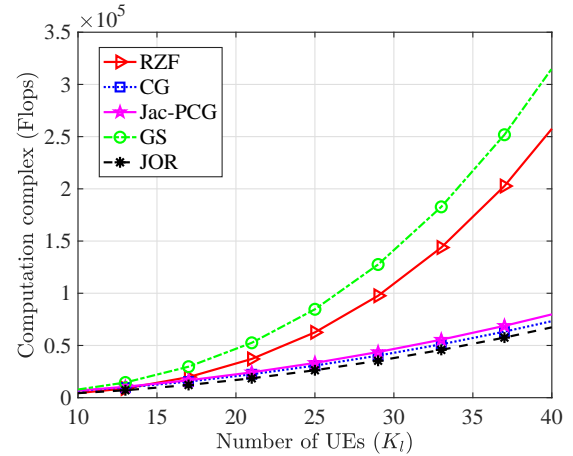}
\caption{Complexity of RZF, GS, JOR,  CG and  Jac-PCG methods against the number of UEs.}
\label{fig:complexity}
\end{figure}

In Fig. \ref{fig:complexity}, we plot the complexity of the RZF, CG, Jac-PCG, GS, and JOR methods against the number of UEs in XL-MIMO systems. The complexity analysis of the investigated methods is provided in Appendix \ref{secA}.
We have found that the CG and Jac-PCG methods have lower complexity than the RZF method of the matrix direct inversion, especially with an extremely large number of UEs. Besides, the complexity of the JOR and CG methods are almost identical and the Jac-PCG method has about $54 \%$ reduction than the traditional RZF method when the number of UEs is $30$. 


\section{Conclusion}
\label{sec:conclusion}
In this paper, we have investigated four iterative matrix inversion methods for XL-MIMO systems, with the goal of fast matrix inversion for RZF precoding.  
Table \ref{Sumary} summarizes the advantages and disadvantages of the iterative methods. Compared to the JOR method, the GS method exhibits a higher SE performance and a faster convergence rate. However, the GS method's computations cannot be parallelized and have high complexity. In contrast, the Jac-PCG method achieves a trade-off between complexity and performance for XL-MIMO systems.

\appendix
	\begin{appendices}
		\subsection{Computational Complexity Analysis of the Investigated Methods}\label{secA}
In this paper, we consider operations on a mixture of real and complex numbers in calculating the computational complexity of the matrix inversion methods examined.
 In order to compare these methods, we need to be more precise than the typical ``order of \begin{math}
\mathcal{O}(\cdot ) 
\end{math}
'' analysis.		

We consider direct matrix inversion of ${\bf P}_{i}$ via Cholesky decomposition. It can be found that (a) the Cholesky decomposition ${\bf P}_{i}={\bf M}_{i}{\bf M}_{i}^{H}$ requires $\frac{4}{3}K^{3}_{l}-K^{2}_{l}+\frac{2}{3}K_{l}+$\begin{math}
\mathcal{O}(K_{l}) 
\end{math}$-1$ flops; (b) finding ${\bf M}_{i}^{-1}$ by forwarding substitution requires $\frac{4}{3}K^{3}_{l}-\frac{1}{3}K_{l}$ flops; and (c) obtaining ${\bf P}_{i}^{-1}=({\bf M}_{i}^{-1})^{H}{\bf M}_{i}^{-1}$ requires $\frac{4}{3}K^{3}_{l}+K^{2}_{l}+\frac{2}{3}K_{l}$ flops. Therefore, in total, direct inversion of ${\bf P}_{i}$ requires $4K^{3}_{l}+K_{l}+$\begin{math}
\mathcal{O}(K_{l}) 
\end{math}$-1$ flops. The GS and JOR methods all calculate ${\bf w}_{i}^{t}={\bf A}_{i}{\bf w}_{i}^{t}+{\bf b}_{i}$,  
which ordinarily requires $8K^{2}_{l}$ flops. However, for GS the first column's elements of ${\bf A}_{i}$ are all zero, this reduces the complexity per iteration for those two methods to $8K^{2}_{l}-8K_{l}$ flops. Initializing GS means solving $({\bf D}_{i}+{\bf L}_{i})[{\bf A}_{i},{\bf b}_{i}]=[-{\bf U}_{i}, {\bf s}_{i}]$ by forward substitution. This results in a total of $4K^{3}_{l}-3K^{2}_{l}+K_{l}$ flops required. For JOR, when calculating ${\bf I}-\omega {\bf D}_{i}^{-1}{\bf P}_{i}$, it takes only $1$ flop to find and set all $K_{l}$ diagonal elements of ${\bf A}_{i}$ equal to the resulting value of $1-\omega$. In total, initialization of JOR requires $2K^{2}_{l}+K_{l}+1$ flops. 

In each iteration of the CG method, the needed steps are: (a) Find ${\bf r}_{i}^{H}{\bf r}_{i}$, which takes $8K_{l}-2$ flops. Calculate ${\bf P}_{i}{\bf m}_{i}$, which 
 requires $8K^{2}_{l}-2K_{l}$ flops, and ${\bf m}_{i}^{H}{\bf P}_{i}{\bf m}_{i}$ requires $8K_{l}-2$. $\alpha _{i} $ requires $8K^{2}_{l}+14K_{l}-4$ flops.
(b) update ${\bf w}_{i}^{t+1}={\bf w}_{i}^{t}+\alpha_{i}^{t} {\bf m}_{i}^{t}$, which requires $8K_{l}$ flops. (c) update ${\bf r}_{i}$, which requires $8K_{l}$ flops. ${\bf b}_{i}^{t}$ requires $8K_{l}-2$ flops. ${\bf m}_{i}^{t}$ requires $8K_{l}$ flops. Thus, the total complexity of the CG method requires $8K^{2}_{l}+46K_{l}-6$ flops. Compared with CG method, Jac-PCG method has one more matrix preprocessing process  
${\bf C}_{i}^{-1} {\bf P}_{i}$, which increases $4K^2_{l}+2K_{l}$ flops.
		
\end{appendices}
\begin{table*}[t!]
  \centering
  \fontsize{9}{13}\selectfont
  \caption{Advantages and Disadvantages of Investigated Iterative Matrix Inversion Methods.}
  \label{Sumary}
   \begin{tabular}{   m{3cm}<{\centering} !{\vrule width0.5pt}  m{1.3 cm}<{\centering} !{\vrule width0.5pt}  m{7.7 cm}<{\centering} !{\vrule width0.5pt} m{5 cm}<{\centering} }

     \hline
       \bf  Method Categories  & \bf Method& \bf Advantages  & \bf  Disadvantages \cr
    
    \hline
    \hline

        \multirowcell{3}{Stationary \\Iterative Methods}  & GS   & \makecell[c]{$\bullet$ Better rate of convergence compared to JOR\\$\bullet$ High SE performance}&\makecell[c]{$\bullet$ High computational complexity\\$\bullet$ Not compatible with parallel \\implementation} \\
               
        \cline{2-4}  & JOR  &\makecell[c]{$\bullet$ Low complexity compared to direct matrix inversion\\$\bullet$ Compatible with parallel/distributed computation} &\makecell[c]{$\bullet$ Slower rate of convergence \\compared to GS}\cr       \hline
       \multirowcell{3}{Non-Stationary \\Iterative Methods}  & CG  &\makecell[c]{$\bullet$ Low complexity compared to GS \\and similar complexity to Jac-PCG}&\makecell{$\bullet$ The SE performance is low} \\
             
        \cline{2-4}   & Jac-PCG &\makecell[c]{$\bullet$ Low complexity compared to GS\\$\bullet$ Better rate of convergence compared to CG\\$\bullet$ Higer SE performance compared to CG}  &\makecell[c]{\\$\bullet$ High computational complexity for \\initialization calculations}  
        
   \cr \hline

    \end{tabular}
  \vspace{0cm}
\end{table*}

\bibliographystyle{IEEEtran}
\bibliography{IEEEabrv,ref}

\end{document}